\documentclass[a4paper,11pt]{article}

\usepackage{amssymb,amsmath}

\newcommand{\ns}{\normalsize}
\newcommand{\be}{\begin{equation}}
\newcommand{\ee}{\end{equation}}
\newcommand{\ba}{\begin{eqnarray}}
\newcommand{\ea}{\end{eqnarray}}

\numberwithin{equation}{section}

\begin{document}


\begin{titlepage}

\vspace{-3cm}

\title{\hfill{\ns hep-th/0412293\\}
   \vskip 1cm
   {\Large Yang-Mills gravity in biconformal space}\\}
   \setcounter{footnote}{0}
\author{
{\ns\large Lara B. Anderson$^1$\footnote{email: anderson@maths.ox.ac.uk}
 \setcounter{footnote}{2}
 and James T. Wheeler$^2$\footnote{email: jwheeler@cc.usu.edu}} \\[1em]
   {\ns\it$^1$Mathematical Institute, University of Oxford}\\
   {\ns 24-29 St.~Giles', Oxford OX1 3LB, UK}\\[1em]
   {\it\ns $^2$Department of Physics}\\
   {\it\ns Utah State University}\\
   {\ns Logan, Utah, 84321, USA}}  
\date{}

\maketitle

\begin{abstract}\noindent 
We write a gravity theory with Yang-Mills type action using the biconformal gauging of the conformal group.
 We show that the resulting biconformal Yang-Mills gravity theories describe 4-dim, scale-invariant
 general relativity in the case of slowly changing fields. In addition, we systematically extend
 arbitrary 4-dim Yang-Mills theories to biconformal space, providing a new arena for studying flat
 space Yang-Mills theories. By applying the biconformal extension to a 4-dim pure Yang-Mills theory
 with conformal symmetry, we establish a 1-1, onto mapping between a set of gravitational gauge
 theories and 4-dim, flat space gauge theories.
\end{abstract}

\thispagestyle{empty}

\end{titlepage}


\section{\smallskip Introduction}

Is it possible to write a Yang-Mills theory of gravity? By such a theory we
mean a Yang-Mills type action,%
\begin{equation}
S_{YM}=\int K_{AB}\mathbf{F}^{A\ast}\mathbf{F}^{B}%
\end{equation}
where the gauge group of $\mathbf{F}^{A}$ contains the Poincar\'{e} group.
While this is a natural action to write for an internal symmetry, the nature
of general relativity as a gauge theory makes $S_{YM}$ unacceptable. The
reason is that viewing GR as a gauge theory requires a caveat: the requirement
of ``soldering,'' i.e., the identification of the solder form with the
cotangent basis. This identification has a disturbing effect on the action
$S_{YM}$ above -- the volume element of the underlying spacetime is expressed
in terms of some of the gauge fields. The result is that the field equations
are not of a purely Yang-Mills type, but also include artificial source terms
built from the curvature. As a result, standard treatments replace $S_{YM}$
above by the Einstein-Hilbert action,%
\begin{equation}
S_{GR}=\int\mathbf{R}^{ab}\mathbf{e}^{c}\mathbf{e}^{d}\varepsilon_{abcd}.
\end{equation}
See, e.g., \cite{Utiyama}-\cite{Held}.

Here, we present a new approach to this question which results from studying a
class of conformal geometries called biconformal spaces. We find gravity
theories with pure Yang-Mills type Lagrangians in $8$-dimensional biconformal
space, thereby constructing Yang-Mills theories of gravity. We refer to these
new theories as biconformal Yang-Mills gravity theories and show that in the
case of slowly changing fields, the theories reduces to general relativity on
a $4$-dimensional sub-manifold.

An interesting further insight into these gravity theories is gained once we
develop a general embedding of $4$-dimensional flat-space Yang-Mills theories
into biconformal space. Applying this embedding to the symmetry of biconformal
Yang-Mills gravity leads to a relationship between pure $4$-dimensional
Yang-Mills solutions and biconformal gravity solutions.

We briefly review here the central concepts of biconformal gravity theories.
The biconformal approach to conformal theory possesses a number of unique
advantages over standard conformal gaugings \cite{Freund1}-\cite{susysohnius}.
In the standard approaches the local symmetry is relativistic similarity group
(Poincar\'{e} plus dilatations). In all of these approaches, difficulties
arise, including the presence of unphysical size changes, the requirement for
an invariant action in $n$ dimensions to be of order $\frac{n}{2}$ in the
curvature, and/or the requirement for auxiliary fields to write a linear action.

An alternative to the standard gaugings was first formulated by Ivanov and
Niederle \cite{IvanovI},\cite{Ivanov}, who created an $8$-dimensional manifold
by gauging the conformal group of a $4$-dimensional spacetime. In their
approach the local symmetry group was taken to be homothetic. Rather than
constraining this space to construct a $4$-dimensional spacetime, they
restricted $4$ of the dimensions as far as possible given the required gauge
freedom. Later these results were extended \cite{JW}, by generalizing to
arbitrary dimensions, $n$, and defining the class of \textit{biconformal
spaces} as the result of the $2n$-dim gauging without imposing constraints. In
that work it was shown that the resulting space possessed symplectic structure
and admitted torsion free spaces consistent with general relativity and
electromagnetism. Further work \cite{WW} provided the most general class of
actions linear in the biconformal curvatures. These models eliminated the
problems listed above, and it was demonstrated that the resulting field
equations lead to the Einstein field equations. The $2n$-dimensional space
constructed by the biconformal gauging is interpreted as a relativistic phase
space of an $n$-dimensional configuration space. This interpretation is
justified by the existence of an integrable symplectic form which guarantees
that the space is a symplectic manifold. The first supersymmetric biconformal
gravity theory was constructed in \cite{AW}.

Returning to the results of this paper, we first investigate the issues
associated with constructing Yang-Mills gravity theories and the successful
model described above. In this discussion we will use unavoidably similar
terminology to discuss several distinct gauge theories. For clarity, we stress
here that below we will define four distinct gauge theories: (1) pure
Yang-Mills theory, (2) Yang-Mills theory coupled to gravity, (3) Yang-Mills
gravity theory and (4) biconformal Yang-Mills gravity theory.

Our second result is a systematic program for embedding $4$-dimensional,
flat-space Yang-Mills theories into biconformal space. We provide a consistent
definition of $8$-dimensional fields in terms of their $4$-dimensional
counterparts. This extension provides a new arena for investigating the
structure and properties of standard Yang-Mills theories.

Finally, we find that while most flat space Yang-Mills theories in
$4$-dimensions are mapped to theories in flat biconformal space, a certain
class of flat Yang-Mills theory can be extended to a gravity theory (with
Yang-Mills Lagrangian) in $8$-dimensions. We establish an exact correspondence
between a non-compact $SU(2,2|N)$ super Yang-Mills theory on flat
$4$-dimensional spacetime and a conformal supergravity theory formulated on a
biconformal space constructed from the same supergroup. The result provides a
previously unknown relationship between a Yang-Mills gauge theory and a
gravity theory and a new class of solutions for biconformal Yang-Mills gravity theories.

In the following section we discuss our notation and the definitions of the
relevant field theories. In particular, we review the definitions of
Yang-Mills theories and Yang-Mills theories coupled to gravity. Then we define
a new class of models, called Yang-Mills gravity theories which are distinct
from both of the theories listed above. In Section 3 we explicitly present a
new class of supergravity models, biconformal Yang-Mills gravity theories.
Developing these models in Section 4, we demonstrate that biconformal
Yang-Mills gravity theories reproduce general relativity on a $4$-dim
submanifold of an $8$-dim space. In Section 5 we develop a formalism for
extending $4$-dimensional, flat space Yang-Mills theories to biconformal
space, beginning with a simple example of a $U(1)$ gauge theory. Section 6
extends the construction to the full non-Abelian case. Further, we show in
Section 7 that the biconformal field equations can be identified with a
non-compact supersymmetric Yang-Mills theory on a flat $4$-dimensional
spacetime. Thus, a correspondence is established between a conformal
supergravity theory and a non-compact Yang-Mills gauge theory. Finally, we
review the results of this work and state conjectures for the structure of
biconformal Yang-Mills supergravity theories.

\section{Definitions and Notation}

We begin with some definitions. Although common usage often restricts
Yang-Mills theories to unitary symmetry, we define a \textit{Yang-Mills
theory} \cite{YangMills},\cite{Weinberg},\cite{BrinkSS} to follow from a
functional of the form
\begin{equation}
S_{YM}=\int K_{AB}\mathbf{F}^{A\ast}\mathbf{F}^{B}%
\end{equation}
(neglecting topological terms) where $\mathbf{F}^{A}$ is the curvature
$2$-form of a principal fiber bundle with \textit{arbitrary} Lie fiber group,
$\mathcal{G}$, and $K_{AB}$ is the Killing metric (here $A,B$ are Lie algebra
indices). The curvatures are related to a connection (or potential),
$\mathbf{A}^{A}$, on the bundle by the Cartan structure equations,
\begin{equation}
\mathbf{F}^{A}=\mathbf{dA}^{A}-\frac{1}{2}c_{BC}{}^{A}\mathbf{A}^{B}%
\mathbf{A}^{C}%
\end{equation}
where the $c_{BC}{}^{A}$ are the structure constants of the group
$\mathcal{G}$. The integral is over a given fixed background manifold. The
symmetry is said to be internal if the Yang-Mills potential is distinct from
the solder form connection on the background manifold. Standard examples
include the $U\left(  1\right)  $ gauge theory of electromagnetism, the
$SU\left(  2\right)  \times U\left(  1\right)  $ electroweak theory, or the
standard model \cite{Utiyama}. The theory is a \textit{flat space Yang-Mills
theory} if there is no curvature of the background space. Then the action
takes the form (in $n$-dim),
\begin{equation}
S_{YM}=\int K_{ab}F_{\alpha\beta}^{a}F_{\mu\nu}^{b}\eta^{\alpha\mu}\eta
^{\beta\nu}d^{n}x \label{Pure YM}%
\end{equation}
where $\alpha,\beta=1...n$.

If the background spacetime is curved, then the theory is a \textit{Yang-Mills
theory coupled to gravity}. To accomplish this, we introduce a general metric,
$g_{\alpha\beta}$ and a volume form $\sqrt{\left|  g\right|  }d^{n}x,$ where
$g=\det\left(  g_{\alpha\beta}\right)  .$ In this case, the gravity action is
added to $S_{YM},$
\begin{align}
S_{YM+G}  &  =\int K_{ab}\mathbf{F}^{a\ast}\mathbf{F}^{b}+\mathbf{R}%
^{ab}\mathbf{e}^{c}\mathbf{e}^{d}\varepsilon_{abcd}\nonumber\\
&  =\int\left(  K_{ab}F_{\alpha\beta}^{a}F_{\mu\nu}^{b}g^{\alpha\mu}%
g^{\beta\nu}+R\right)  \sqrt{|g|}d^{n}x \label{YM coupled to grav}%
\end{align}
Variation of the metric then leads to $n$-dim general relativity with the
energy-momentum tensor associated with $F_{\alpha\beta}^{a}$ as gravitational
source. The Yang-Mills field $F_{\alpha\beta}^{a}$ evolves in the curved
background described by $g_{\alpha\beta}.$ Note that the Hodge dual
automatically brings in couplings to the curved metric. In both flat space
Yang-Mills and Yang-Mills coupled to gravity, the symmetry is internal.

Before we define Yang-Mills gravity, we require a digression to characterize
gravitational gauge theories. Despite our very general definition of
Yang-Mills gauge theory, there are other types of gauge theory. One simple
variation is to choose any other group invariant action to replace the
Yang-Mills action. But there is a deeper difference that arises when we
consider a gravitational gauge theory. Suppose we select the Poincar\'{e}
group as our symmetry group. In this case, one of the potentials of the
Yang-Mills field, the gauge field of translations, $\mathbf{A}^{a}%
=\mathbf{e}^{a}=\mathbf{d}x^{\alpha}e_{\alpha}^{\quad a},$ called the solder
form, also describes the space in which the force acts via the relation
\begin{equation}
g_{\alpha\beta}=e_{\alpha}{}^{a}e_{\beta}{}^{b}\eta_{ab}
\label{Metric to solder form}%
\end{equation}
where $\eta_{ab}$ is the Minkowski metric \cite{Kobayashi}. This means that
the symmetry is no longer internal. The identification of the solder form
$\mathbf{e}^{a}$ as basis forms for the cotangent space (``soldering'') breaks
translational invariance. The degrees of freedom associated with translations
are not lost, however, but reappear as general coordinate invariance. The
final spacetime does not have full Poincar\'{e} invariance, rather, the local
symmetry is Lorentz only \cite{Kibble1}, \cite{Isham}, \cite{Held}.

One drawback of this ``soldering'' approach to gravitational gauge theory is
the redundancy of introducing both a $10$-dim symmetry group and a separate
$4$-dim background space, and then identifying four of the dimensions of the
symmetry group with those of the background space. It is tempting to think
that the introduction of an independent background space might be avoided.
This hope is realized by the group quotient or group manifold method
\cite{Neeman},\cite{ReggeN},\cite{Utiyama}, in which the local symmetry and
the manifold are simultaneously constructed from an enveloping symmetry. For
the Poincar\'{e} example, we may construct the base manifold by taking the
quotient of the Poincar\'{e} group by the Lorentz group. The result is a
principal fiber bundle with Lorentz fibers over a $4$-dim (or $n$-dim)
manifold. The Cartan connection on this manifold is generalized by introducing
horizontal curvature $2$-forms into the Maurer-Cartan equations. This
procedure gives not only local Lorentz symmetry as developed by Utiyama
\cite{Utiyama}, but also the solder form as employed by Kibble \cite{Kibble1},
without the separate introduction of a background space. The model always
retains the overall dimension of the original group, with the dimensions
partitioned between the local symmetry and the manifold. One can then write an
action using the curvatures, solder form and invariant tensors in a way
respecting the remaining local Lorentz symmetry. One advantage of these
group-theoretic methods is that the fields are determined by the symmetry.
This leaves a great deal of freedom in choosing the action, and suggests a way
to write a theory that simultaneously looks like Yang-Mills and gravity.

We may now define a \textit{Yang-Mills gravity theory} to be a Yang-Mills
theory with a symmetry group containing the Poincar\'{e} group (Poincar\'{e},
Weyl, conformal, their covering groups, and their supersymmetric extensions)
together with a Yang-Mills action
\begin{align}
S_{YMG}  &  =\int K_{AB}\mathbf{\Omega}^{A\ast}\mathbf{\Omega}^{B}\nonumber\\
&  =\int K_{AB}\Omega_{\alpha\beta}^{A}\Omega_{\mu\nu}^{B}g^{\alpha\mu
}g^{\beta\nu}ed^{n}x \label{YMGrav}%
\end{align}
where $e=\sqrt{\left|  g\right|  }$.

In light of the preceding paragraphs, we avoid introducing a separate
background geometry. Thus, the metric tensor, $g_{\alpha\beta}$ is constructed
from translational gauge fields, $\mathbf{e}^{\Sigma},$ $\Sigma=1,\ldots,n,$
according to eq.(\ref{Metric to solder form}). As a result, some of the
original symmetry is broken, with the corresponding transformations replaced
by diffeomorphisms. The field equations differ from a pure Yang-Mills theory,
while at the same time differing from general relativity. In particular, the
field equations arising from variation of the translational gauge fields will
be of the form
\[
D^{\alpha}\Omega_{\alpha\beta}^{\Sigma}=T_{\quad\beta}^{\Sigma}%
\]
where the derivative is covariant with respect to the remaining local symmetry
and general coordinate transformations. The appearance of an energy-momentum
tensor, $T_{\quad\beta}^{\Sigma}$, which makes the theory differ from pure
Yang Mills, is due to the variation of the extra metric terms, $g^{\alpha\mu
}g^{\beta\nu}e$. At the same time, the derivative of curvature on the left
gives a non-standard gravity theory. Note that the difference between
eq.(\ref{YMGrav}) and the pure Yang-Mills theory of eq.(\ref{Pure YM}) lies in
the dual. Because the solder form that determines the volume form is one of
the gauge fields, the dual introduces non-Yang-Mills coupling, $g^{\alpha\mu
}g^{\beta\nu}e$.

In addition to the difficulty reproducing either pure Yang-Mills or standard
gravity, the Poincar\'{e} group encounters an immediate problem with its lack
of a Killing metric. Writing the action using the degenerate Killing form
leads to vanishing of the energy-momentum tensor constructed from the
Riemannian curvature (i.e., $R_{bce}^{a}R_{adf}^{b}g^{ef}-g_{cd}R_{beg}%
^{a}R_{afh}^{b}g^{ef}g^{gh})$ -- a highly artificial constraint. The other
logical symmetry choices of the Weyl group $%
\genfrac{(}{)}{}{}{ISO\left(  3,1\right)  \times R^{+}}{SO\left(  3,1\right)
\times R^{+}}%
$ and conformal group with standard gauging $%
\genfrac{(}{)}{}{}{SO\left(  4,2\right)  }{ISO\left(  3,1\right)  \times
R^{+}}%
$ lead to higher order field equations and non-Yang-Mills couplings. However,
as we shall argue below, there exists a Yang-Mills gravity theory which solves
the coupling problem.

\textit{Biconformal gauging} of $SU(2,2)$ (or of $SU(2,2|$ $N)$ ) is unique in
solving some or all of these problems. By taking the quotient of the conformal
group by the homogeneous homothetic group (Lorentz plus dilatations) instead
of the inhomogeneous homothetic group (Poincar\'{e} plus dilatations),
biconformal gauging produces a symplectic base manifold with local Lorentz and
dilatational symmetry. The symplectic manifold \textit{has a natural,
dimensionless volume element which may be written without using the gauge
fields}. This allows the gravitational symmetry to remain internal. Moreover,
the reduction of local symmetry reduces the conformally-covariant divergence
of the Yang-Mills field equation to a Weyl-covariant divergence together with
terms algebraic in the curvatures. These algebraic terms can reproduce general
relativity in a suitable limit. These improvements are the topic of the
remainder of next two sections.

\section{Biconformal Yang-Mills Gravity}

By \textit{biconformal Yang-Mills gravity}, we mean a Yang-Mills gravity
theory of the conformal group, its covering group or its supersymmetric
extension in which the gravitational gauging is accomplished by writing the
theory on the $2n$-bosonic dimensional quotient manifold of the
(super)conformal group by its Lorentz plus dilatational sub-(super)group.
These gaugings are described more fully elsewhere \cite{JW},\cite{WW}%
,\cite{AW} (and further properties of biconformal space are examined in
\cite{GaugeNewton},\cite{QM}). For simplicity, we treat only the $n=4$ case.

Biconformal Yang-Mills gravity makes use of the fact that biconformal spaces
have coordinates which are naturally adapted to a symplectic manifold. Half
the biconformal coordinates arise from translations, while the other half
arise from special conformal transformations (co-translations). Since we may
always choose coordinates of these two types, biconformal spaces have almost
symplectic structure
\[
\mathbf{\Theta}=\mathbf{d}x^{\alpha}\mathbf{d}y_{\alpha}%
\]
The structure is \textit{almost} symplectic because we are guaranteed these
coordinates on each chart of the manifold. The structure may or may not be
integrable globally. However, all known classes of biconformal solutions do
give rise to a global symplectic form. The symplectic form $\mathbf{\Theta}$
enables us to write the volume form as
\[
\mathbf{\Phi}=\left(  \mathbf{d}x^{\alpha}\mathbf{d}y_{\alpha}\right)  ^{4}%
\]
Taking the dual relative to this volume form does not introduce spurious gauge
fields. For a $2$-form
\begin{equation}
\mathbf{\Omega}^{A}=\frac{1}{2}\Omega_{\quad\mu\nu}^{A}\mathbf{d}x^{\mu
}\,\mathbf{d}x^{\nu}+\Omega_{\quad\nu}^{A\mu}\mathbf{d}y_{\mu}\,\mathbf{d}%
x^{\nu}+\frac{1}{2}\Omega^{A\mu\nu}\mathbf{d}y_{\mu}\,\mathbf{d}y_{\nu}%
\end{equation}
($A,B=1,\ldots,15$) the dual is simply
\begin{align}
^{\ast}\mathbf{\Omega}^{B}  &  =\frac{1}{2\cdot4!\cdot2!}\Omega_{\quad
\mu^{\prime}\nu^{\prime}}^{B}\varepsilon^{\mu^{\prime}\nu^{\prime}\alpha\beta
}\mathbf{d}y_{\alpha}\mathbf{d}y_{\beta}\varepsilon_{\rho\sigma\gamma\delta
}\mathbf{d}x^{\rho}\,\mathbf{d}x^{\sigma}\mathbf{d}x^{\gamma}\,\mathbf{d}%
x^{\delta}\nonumber\\
&  +\frac{1}{3!\cdot3!}\Omega_{\quad\nu^{\prime}}^{B\mu^{\prime}}%
\varepsilon^{\nu^{\prime}\alpha\beta\rho}\mathbf{d}y_{\alpha}\mathbf{d}%
y_{\beta}\mathbf{d}y_{\rho}\varepsilon_{\mu^{\prime}\sigma\gamma\delta
}\,\mathbf{d}x^{\sigma}\mathbf{d}x^{\gamma}\,\mathbf{d}x^{\delta}\nonumber\\
&  +\frac{1}{2\cdot4!\cdot2!}\Omega^{B\mu^{\prime}\nu^{\prime}}\varepsilon
^{\alpha\beta\rho\sigma}\mathbf{d}y_{\alpha}\mathbf{d}y_{\beta}\mathbf{d}%
y_{\rho}\mathbf{d}y_{\sigma}\varepsilon_{\mu^{\prime}\nu^{\prime}\gamma\delta
}\mathbf{d}x^{\gamma}\,\mathbf{d}x^{\delta}%
\end{align}
and the action is of pure Yang-Mills form,%

\begin{align}
S  &  =\frac{1}{4}\int K_{AB}\mathbf{\Omega}^{A\ast}\mathbf{\Omega}%
^{B}\nonumber\\
&  =\frac{1}{4}\int K_{AB}\left(  \Omega_{\quad\mu\nu}^{A}\Omega^{B\mu\nu
}-\Omega_{\quad\nu}^{A\mu}\Omega_{\quad\mu}^{B\nu}\right)  \mathbf{\Phi}
\label{BYMG}%
\end{align}
There are no gauge fields in the volume element, $\mathbf{\Phi,}$ nor is the
metric required to form the various contractions. Notice that the
coordinate-based, metric-free action does not exclude the possibility of an
equivalent orthonormal basis.

We define the anti-symmetric symbol in biconformal space to be,
\[
\varepsilon_{M_{1}\cdots M_{8}}=\varepsilon_{\lbrack M_{1}\cdots M_{4}}%
\delta_{M_{5}}^{4+N_{1}}\cdots\delta_{M_{8}]}^{4+N_{4}}\varepsilon
_{N_{1}\cdots N_{4}}%
\]
where $M_{i}=1,2,\ldots,8$ and $N_{i}=1,2,\ldots,4$. It is important to
recognize that in this geometry, the anti-symmetric symbol is already a
tensor, not a tensor density. To see this, recall that in spacetime integrals,
we require the tensor
\[
e_{\mu\nu\gamma\delta}=\sqrt{\left|  g\right|  }\varepsilon_{\mu\nu
\gamma\delta}%
\]
in writing the volume form. But, as indicated above, the full biconformal
Levi-Civita tensor reduces to a product of $4$-dim Levi-Civita symbols,
$\varepsilon^{\alpha\beta\rho\sigma}\varepsilon_{\mu\nu\gamma\delta}.$ Each of
these forms is separately a tensor density, but since they have opposite
weights the product is a tensor without the addition of any metric dependent
terms. That is,
\[
e^{\alpha\beta\rho\sigma}e_{\mu\nu\gamma\delta}=\frac{1}{\sqrt{\left|
g\right|  }}\varepsilon^{\alpha\beta\rho\sigma}\sqrt{\left|  g\right|
}\varepsilon_{\mu\nu\gamma\delta}=\varepsilon^{\alpha\beta\rho\sigma
}\varepsilon_{\mu\nu\gamma\delta}%
\]

Moreover, because of the ``doubled'' volume form, the components of the dual
arising from
\[
\Omega^{B}{}_{\mu\nu}%
\]
may be written as
\[
\Omega^{B}{}_{\mu\nu}\varepsilon^{\mu\nu\alpha\beta}\varepsilon_{\rho
\sigma\gamma\delta}%
\]
so no metric is required. As a result, $\Omega^{A}{}_{\mu\nu}$ contracts with
$\Omega^{B\mu\nu}$ and $\Omega_{\quad\nu}^{A\mu}$ with itself, again requiring
no metric. This means that the usual metric contribution
\[
g^{\alpha\mu}g^{\beta\nu}\sqrt{\left|  g\right|  }%
\]
to Yang-Mills gravity is absent in biconformal Yang-Mills gravity.

These results carry through when we include supersymmetry. When we gauge the
supergroup $SU\left(  2,2|N\right)  $ as a biconformal Yang-Mills gravity
theory, taking the group quotient, $%
\genfrac{(}{)}{}{}{SU\left(  2,2|N\right)  }{R^{+}\times SL\left(  2,C\right)
\times SU\left(  N\right)  }%
,$ we also do not require any extra factors involving the gauge fields. In
this case, the superspace volume form,
\[
\mathbf{\Phi}=\left(  \mathbf{d}x^{\alpha}\mathbf{d}y_{\alpha}\right)
^{4}\left(  \theta_{\alpha}^{A}\bar{\theta}_{A}^{\alpha}\right)  ^{4N}%
\]
is also metric-free, since the fermionic variables also occur in dimensionally
conjugate pairs.

There is also the question of coordinate invariance. How can the action be
written without the usual metric determinants, yet still describe
diffeomorphism invariant gravity? The answer lies in the symplectic structure.
The volume form $\mathbf{\Phi}=\left(  \mathbf{d}x^{\alpha}\mathbf{d}%
y_{\alpha}\right)  ^{4}$ is invariant under $8$-dim symplectic
transformations, and these include general coordinate transformations of the
$4$-dim configuration space. When we perform the reduction to the
configuration space as described above, setting $y_{\alpha}=y_{\alpha}\left(
x^{\beta}\right)  ,$ we must either introduce a connection or continue to
transform $y_{\alpha}$ inversely to $x^{\alpha}$.

Notice that conformal supergravity theories exist in $4$, $6$ and $10$ but not
in $8$ dimensions. The $8$-dim biconformal theory uses the superconformal
group associated with $4$-dim spacetime. It still describes $4$-dim gravity.

For simplicity and clarity, we restrict attention to the bosonic conformal
sector of the symmetry group for the remainder of the paper. There appears to
be no obstacle to writing the theory for the full supergroup. In the next
section we examine the gravity equations that arise from the Yang-Mills
action. Remarkably, they still lead to general relativity when the Weyl vector
vanishes or in general for weak, slowly changing fields.

\section{Reduction to general relativity}

In this section we produce our next central result: the reduction of
biconformal Yang-Mills gravity theory to general relativity in a certain
limit. The field equations for the $SU(2,2)$ part of the bosonic sector are
the usual Yang-Mills field equations,
\begin{align*}
D_{\alpha}\Omega^{B\alpha\beta}-D^{\alpha}\Omega_{\quad\alpha}^{B\beta}  &
=0\\
D^{\alpha}\Omega_{\quad\alpha\beta}^{B}+D_{\alpha}\Omega_{\quad\beta}%
^{B\alpha}  &  =0
\end{align*}
where the derivatives are conformally covariant with respect to the Latin
(group) indices only:
\begin{align*}
D_{\mu}^{Conf}\Omega^{B\alpha\beta}  &  =\partial_{\mu}\Omega^{B\alpha\beta
}+\Omega^{C\alpha\beta}\omega_{C\mu}^{B}\\
&  =\partial_{\mu}\Omega^{B\alpha\beta}-\frac{1}{2}c_{CD}{}^{B}\Omega
^{C\alpha\beta}\omega_{\mu}^{D}%
\end{align*}
where $c_{CD}{}^{B}$ are the structure constant of the conformal group.
Normally, this entire expression comprises a single tensor. However, the
biconformal quotient breaks the conformal invariance to homothetic (Weyl)
invariance. As a result, the gauge fields of translations $\mathbf{\omega}%
^{a}$ (``solder form'') and the gauge fields of special conformal
transformations, $\mathbf{\omega}_{a}$ (``co-solder form'') become tensors.
The conformally covariant derivative then reduces to a Weyl covariant
derivative plus tensor terms linear in the curvatures. Specifically, we have
\begin{align*}
D_{S}^{Conf}\Omega_{bMN}^{a}  &  =\partial_{S}\Omega_{bMN}^{a}-\omega_{bS}%
^{c}\Omega_{cMN}^{a}+\Omega_{bMN}^{c}\omega_{cS}^{a}+4\Delta_{db}^{ac}\left(
\Omega_{cMN}\omega_{S}^{d}-\omega_{cS}\Omega_{MN}^{d}\right) \\
&  =D_{S}^{Weyl}\Omega_{bMN}^{a}+4\Delta_{db}^{ac}\left(  \Omega_{cMN}%
\omega_{S}^{d}-\omega_{cS}\Omega_{MN}^{d}\right) \\
D_{S}^{Conf}\Omega_{MN}^{a}  &  =\partial_{S}\Omega^{a}+\Omega_{MN}^{c}%
\omega_{cS}^{a}-\omega_{S}\Omega_{MN}^{a}+\Omega_{MN}\omega_{S}^{a}-\omega
_{S}^{c}\Omega_{cMN}^{a}\\
&  =D_{S}^{Weyl}\Omega_{MN}^{a}+\Omega_{MN}\omega_{S}^{a}-\omega_{S}^{c}%
\Omega_{cMN}^{a}\\
D_{S}^{Conf}\Omega_{aMN}  &  =\partial_{S}\Omega_{aMN}-\omega_{aS}^{c}%
\Omega_{cMN}+\Omega_{aMN}\omega_{S}+\Omega_{aMN}^{b}\omega_{bS}-\omega
_{aS}\Omega_{MN}\\
&  =D_{S}^{Weyl}\Omega_{aMN}+\Omega_{aMN}^{b}\omega_{bS}-\omega_{aS}%
\Omega_{MN}\\
D_{S}^{Conf}\Omega_{MN}  &  =\partial_{S}\Omega_{MN}+\Omega_{MN}^{c}%
\omega_{cS}-\Omega_{cMN}\omega_{S}^{c}\\
&  =D_{S}^{Weyl}\Omega_{MN}+\Omega_{MN}^{c}\omega_{cS}-\omega_{S}^{c}%
\Omega_{cMN}%
\end{align*}
where $a,b=1...4$ and $\mathbf{\Omega}_{b}^{a}$, $\mathbf{\Omega}_{a}$,
$\mathbf{\Omega}^{a}$, $\mathbf{\Omega}$ are the curvature, torsion,
co-torsion, and dilatation, respectively. This leads to the field equations
\begin{align*}
0  &  =D_{\beta}^{Weyl}\Omega_{b\alpha}^{a\beta}+D_{Weyl}^{\beta}%
\Omega_{b\beta\alpha}^{a}+4\Delta_{db}^{ac}\left(  \Omega_{c\quad\alpha
}^{\quad\beta}\omega_{\quad\beta}^{d}-\omega_{c}^{\quad\beta}\Omega
_{\quad\beta\alpha}^{d}\right) \\
&  +4\Delta_{db}^{ac}\left(  \Omega_{c\beta\alpha}^{\quad}\omega^{d\beta
}-\omega_{c\beta}\Omega_{\quad\alpha}^{d\beta}\right) \\
0  &  =D_{\beta}^{Weyl}\Omega_{\quad\alpha}^{a\beta}+D_{Weyl}^{\beta}%
\Omega_{\quad\beta\alpha}^{a}+\Omega_{\beta\alpha}\omega^{a\beta}%
-\omega_{\quad\beta}^{c}\Omega_{c\quad\alpha}^{a\beta}\\
&  +\Omega_{\alpha}^{\beta}\omega_{\quad\beta}^{a}-\omega^{c\beta}%
\Omega_{c\beta\alpha}^{a}\\
0  &  =D_{\beta}^{Weyl}\Omega_{a\quad\alpha}^{\quad\beta}+D_{Weyl}^{\beta
}\Omega_{a\beta\alpha}+\Omega_{a\beta\alpha}^{b}\omega_{b}^{\quad\beta}%
-\omega_{a\beta}\Omega_{\quad\alpha}^{\beta}\\
&  +\Omega_{a\quad\alpha}^{b\beta}\omega_{b\beta}-\omega_{a}^{\quad\beta
}\Omega_{\beta\alpha}\\
0  &  =D_{\beta}^{Weyl}\Omega_{\quad\alpha}^{\beta}+D_{Weyl}^{\beta}%
\Omega_{\beta\alpha}+\Omega_{\quad\beta\alpha}^{c}\omega_{c}^{\quad\beta
}-\omega_{\quad\beta}^{c}\Omega_{c\quad\alpha}^{\quad\beta}\\
&  +\Omega_{\quad\alpha}^{c\beta}\omega_{c\beta}-\omega^{c\beta}\Omega
_{c\beta\alpha}%
\end{align*}
and
\begin{align*}
0  &  =D_{\beta}^{Weyl}\Omega_{b}^{a\beta\alpha}+D_{Weyl}^{\beta}%
\Omega_{b\beta}^{a\quad\alpha}+4\Delta_{db}^{ac}\left(  \Omega_{c}^{\quad
\beta\alpha}\omega_{\quad\beta}^{d}-\omega_{c}^{\quad\beta}\Omega_{\quad\beta
}^{d\quad\alpha}\right) \\
&  +4\Delta_{db}^{ac}\left(  \Omega_{c\beta}^{\quad\quad\alpha}\omega^{d\beta
}-\omega_{c\beta}\Omega^{d\beta\alpha}\right) \\
0  &  =D_{\beta}^{Weyl}\Omega_{\quad\beta}^{a\quad\quad\alpha}+D_{Weyl}%
^{\beta}\Omega_{\quad\beta}^{a\quad\alpha}+\Omega_{\beta}^{\quad\alpha}%
\omega^{a\beta}-\omega_{\quad\beta}^{c}\Omega_{c}^{a\beta\alpha}\\
&  +\Omega^{\beta\alpha}\omega_{\quad\beta}^{a}-\omega^{c\beta}\Omega_{c\beta
}^{a\quad\alpha}\\
0  &  =D_{\beta}^{Weyl}\Omega_{a}^{\quad\beta\alpha}+D_{Weyl}^{\beta}%
\Omega_{a\beta}^{\quad\alpha}+\Omega_{a\beta}^{b\quad\alpha}\omega_{b}%
^{\quad\beta}-\omega_{a\beta}\Omega^{\beta\alpha}\\
&  +\Omega_{a}^{b\beta\alpha}\omega_{b\beta}-\omega_{a}^{\quad\beta}%
\Omega_{\beta}^{\quad\alpha}\\
0  &  =D_{\beta}^{Weyl}\Omega^{\beta\alpha}+D_{Weyl}^{\beta}\Omega_{\beta
}^{\quad\alpha}+\Omega_{\quad\beta}^{c\quad\alpha}\omega_{c}^{\quad\beta
}-\omega_{\quad\beta}^{c}\Omega_{c}^{\quad\beta\alpha}\\
&  +\Omega^{c\beta\alpha}\omega_{c\beta}-\omega^{c\beta}\Omega_{c\beta}%
^{\quad\alpha}%
\end{align*}
Since the solder and co-solder forms, ($\mathbf{\omega}_{a}=\omega_{a\beta
}\mathbf{d}x^{\beta}+\omega_{a}^{\quad\beta}\mathbf{d}y_{\beta}$,
$\mathbf{\omega}^{a}=\omega_{\quad\beta}^{a}\mathbf{d}x^{\beta}+\omega
^{a\beta}\mathbf{d}y_{\beta}$) are tensors, each field equation has an
algebraic piece in addition to the usual divergence terms of a Yang-Mills theory.

In torsion-free biconformal spaces, the second equation becomes purely
algebraic:
\begin{equation}
0=\Omega_{\beta\alpha}\omega^{a\beta}-\omega_{\quad\beta}^{c}\Omega
_{c\quad\alpha}^{a\beta}+\Omega_{\alpha}^{\beta}\omega_{\quad\beta}^{a}%
-\omega^{c\beta}\Omega_{c\beta\alpha}^{a} \label{Einstein}%
\end{equation}
This equation is the biconformal equivalent of the Einstein equation. In
biconformal spaces with actions linear in the curvature, torsion-free
solutions are known to descend to general relativity on a submanifold
\cite{JW},\cite{WW}.

One class of exact gravity solutions to the Yang-Mills gravity solutions is
known \cite{Spencer}. It is found that when the dilatational curvature and
torsion vanish, the field equations of the linear-curvature theory
\cite{JW},\cite{WW} and the field equations of the full Yang-Mills gravity
theory are simultaneously satisfied. The common solution is completely
characterized by solutions to the vacuum Einstein equation in $4$-dimensions.

Whether the torsion vanishes or not, there is a range of solutions to these
equations when the Weyl-covariant derivative of the curvatures is small. In
this regime, the field equations reduce to the algebraic, curvature-linear
equations of familiar gravity theories. For example, consider a solution which
reduces under suitable conditions to the Schwarzschild solution. Choosing a
local frame in which the connection vanishes at a point, the curvatures are
inversely proportional to the square of the radial coordinate, while the
derivatives of the curvatures vary as the inverse cube of the radial
coordinate. Since the only relevant constant with units of length is the
Schwarzschild radius, the condition for the derivative terms to be larger than
the curvature terms must be
\[
\frac{1}{r^{3}}\gtrsim\frac{1}{r_{0}r^{2}}%
\]
or simply
\[
r\lesssim r_{0}%
\]
It is therefore only within the horizon that the solutions differ
significantly from a solution to the algebraic equations.

In any region where we may neglect derivatives of the curvature, the field
equations reduce to:
\begin{align*}
0  &  =4\Delta_{db}^{ac}\left(  \Omega_{c\quad\alpha}^{\quad\beta}%
\omega_{\quad\beta}^{d}+\Omega_{c\beta\alpha}^{\quad}\omega^{d\beta}%
-\omega_{c}^{\quad\beta}\Omega_{\quad\beta\alpha}^{d}-\omega_{c\beta}%
\Omega_{\quad\alpha}^{d\beta}\right) \\
0  &  =\Omega_{\beta\alpha}\omega^{a\beta}+\Omega_{\alpha}^{\beta}%
\omega_{\quad\beta}^{a}-\omega_{\quad\beta}^{c}\Omega_{c\quad\alpha}^{a\beta
}-\omega^{c\beta}\Omega_{c\beta\alpha}^{a}\\
0  &  =\Omega_{a\beta\alpha}^{b}\omega_{b}^{\quad\beta}+\Omega_{a\quad\alpha
}^{b\beta}\omega_{b\beta}-\omega_{a\beta}\Omega_{\quad\alpha}^{\beta}%
-\omega_{a}^{\quad\beta}\Omega_{\beta\alpha}\\
0  &  =\Omega_{\quad\beta\alpha}^{c}\omega_{c}^{\quad\beta}+\Omega
_{\quad\alpha}^{c\beta}\omega_{c\beta}-\omega_{\quad\beta}^{c}\Omega
_{c\quad\alpha}^{\quad\beta}-\omega^{c\beta}\Omega_{c\beta\alpha}\\
0  &  =4\Delta_{db}^{ac}\left(  \Omega_{c}^{\quad\beta\alpha}\omega
_{\quad\beta}^{d}+4\Delta_{db}^{ac}\Omega_{c\beta}^{\quad\quad\alpha}%
\omega^{d\beta}-\omega_{c}^{\quad\beta}\Omega_{\quad\beta}^{d\quad\quad\alpha
}-\omega_{c\beta}\Omega^{d\beta\alpha}\right) \\
0  &  =\Omega_{\beta}^{\quad\alpha}\omega^{a\beta}+\Omega^{\beta\alpha}%
\omega_{\quad\beta}^{a}-\omega_{\quad\beta}^{c}\Omega_{c}^{a\beta\alpha
}-\omega^{c\beta}\Omega_{c\beta}^{a\quad\alpha}\\
0  &  =\Omega_{a\beta}^{b\quad\quad\alpha}\omega_{b}^{\quad\beta}+\Omega
_{a}^{b\beta\alpha}\omega_{b\beta}-\omega_{a\beta}\Omega^{\beta\alpha}%
-\omega_{a}^{\quad\beta}\Omega_{\beta}^{\quad\alpha}\\
0  &  =\Omega_{\quad\beta}^{c\quad\quad\alpha}\omega_{c}^{\quad\beta}%
+\Omega^{c\beta\alpha}\omega_{c\beta}-\omega_{\quad\beta}^{c}\Omega_{c}%
^{\quad\beta\alpha}-\omega^{c\beta}\Omega_{c\beta}^{\quad\quad\alpha}%
\end{align*}
\newline These equations are satisfied by the known torsion-free solutions,
which describe general relativity on a $4$-dim submanifold.

\section{U(1) Yang Mills in biconformal space}

We will now pause in our development of a gravity theory to consider the
structure of a non-gravitational gauge theory on biconformal space. In
particular, in this section we will show that any $4$-dim pure Yang-Mills
theory can be consistently written in biconformal space. We establish an
invertible map between the $4$-dim and $8$-dim theories and show that the
$8$-dim field equations are satisfied if and only if the $4$-dim equations are
satisfied. We provide the construction here in detail for $U\left(  1\right)
$. The non-Abelian case is discussed in the next section.

Suppose we have a $U(1)$ gauge theory in $4$-dim, with structure equation,
field equation and conservation law, given by%
\begin{align}
\mathbf{A}  &  =A_{a}\left(  x\right)  \mathbf{d}x^{a}\\
\mathbf{F}  &  =\mathbf{dA}\\
\mathbf{dF}  &  =\mathbf{d}^{2}\mathbf{A}=0\\
^{\ast}\mathbf{d}^{\ast}\mathbf{F}  &  =\mathbf{J}\\
^{\ast}\mathbf{d}^{\ast}\mathbf{J}  &  =0
\end{align}
We wish to extend this to biconformal space. First, we show that the
co-tangent bundle extends to a biconformal space. Let $\left(  x^{a}%
,y_{b}\right)  $ be coordinates on the bundle, which we treat as an $8$-dim
manifold. Extend the bundle by $7$-dim fibers, with each fiber isomorphic to
the spacetime homothetic group. We may then define a connection on the
resulting $15$-dim bundle,
\begin{align}
\mathbf{\omega}  &  =-y_{a}\mathbf{d}x\nonumber\\
\mathbf{\omega}^{a}  &  =\mathbf{d}x^{a}\nonumber\\
\mathbf{\omega}_{a}  &  =\mathbf{d}y_{a}-\alpha\left(  y_{a}y_{b}-\frac{1}%
{2}y^{2}\eta_{ab}\right)  \mathbf{d}x^{b}\nonumber\\
\mathbf{\omega}_{b}^{a}  &  =\left(  y_{b}\mathbf{d}x^{a}-y^{a}\mathbf{d}%
x_{b}\right)
\end{align}
This connection satisfies the Maurer-Cartan structure equations for the
conformal group,
\begin{align}
\mathbf{d\omega}  &  =\mathbf{\omega}^{a}\mathbf{\omega}_{a}\nonumber\\
\mathbf{d\omega}^{a}  &  =\mathbf{\omega}^{b}\mathbf{\omega}_{b}%
^{a}+\mathbf{\omega\omega}^{a}\nonumber\\
\mathbf{d\omega}_{a}  &  =\mathbf{\omega}_{a}^{b}\mathbf{\omega}%
_{b}+\mathbf{\omega}_{a}\mathbf{\omega}\nonumber\\
\mathbf{d\omega}_{b}^{a}  &  =\mathbf{\omega}_{b}^{c}\mathbf{\omega}_{c}%
^{a}+2\Delta_{db}^{ac}\mathbf{\omega}_{c}\mathbf{\omega}^{d}%
\end{align}
Because the bundle is homothetic over an $8$-dim base manifold, it is a
biconformal space.

Next we extend the field theory to the biconformal space. Let%
\begin{align}
\mathcal{A}  &  =\mathbf{A}+\frac{1}{2}F^{ab}y_{a}\mathbf{d}y_{b}\nonumber\\
\mathcal{F}  &  =\mathbf{d}\mathcal{A}\nonumber\\
&  =\mathbf{F}-\frac{1}{2}y_{a}F_{\quad,c}^{ab}\mathbf{d}y_{b}\mathbf{d}%
x^{c}\mathbf{+}\frac{1}{2}F^{ab}\mathbf{d}y_{a}\mathbf{d}y_{b}%
\end{align}
It follows that%
\begin{align}
\mathbf{d}\mathcal{F}  &  =\mathbf{d}^{2}\mathcal{A}=0\nonumber\\
^{\ast}\mathbf{d}^{\ast}\mathcal{F}  &  =-\frac{1}{2}y_{a}F_{\quad,cb}%
^{ab}\mathbf{d}x^{c}+\frac{1}{2}F_{\quad,a}^{ab}\mathbf{d}y_{b}+F_{\quad
,a}^{ab}\mathbf{d}y_{b}%
\end{align}
Now from the original $4$-dim Yang-Mills theory we have%
\begin{equation}
F_{\quad,a}^{ab}=J^{b}%
\end{equation}
and it follows that%
\begin{equation}
^{\ast}\mathbf{d}^{\ast}\mathcal{F}=\frac{3}{2}J^{b}\mathbf{d}y_{b}%
-\frac{1}{2}y_{a}J_{\quad,c}^{a}\mathbf{d}x^{c}%
\end{equation}
Since the right-hand side is constructed purely from the $4$-dim current, we
may define%

\begin{equation}
\mathcal{J}=\frac{3}{2}J^{a}\mathbf{d}y_{a}-\frac{1}{2}y_{a}J_{\quad,b}%
^{a}\mathbf{d}x^{b}%
\end{equation}
to write the $8$-dim biconformal equation%
\begin{equation}
^{\ast}\mathbf{d}^{\ast}\mathcal{F}=\mathcal{J}.
\end{equation}
This completes the biconformal equations governing $\mathcal{A}$,
$\mathcal{F}$ and $\mathcal{J}$. Conservation of $\mathcal{J}$ follows from
$^{\ast}\mathbf{d}^{\ast}{}^{\ast}\mathbf{d}^{\ast}=0$, but is dependent upon
the conservation of the original $4$-current, since direct calculation shows
that%
\begin{align}
^{\ast}\mathbf{d}^{\ast}\mathcal{J}  &  =^{\ast}\mathbf{d}^{\ast}\left(
\frac{3}{2}J^{a}\mathbf{d}y_{a}-\frac{1}{2}y_{a}J_{\quad,b}^{a}\mathbf{d}%
x^{b}\right) \nonumber\\
&  =J_{\quad,a}^{a}%
\end{align}

It is now immediate that for any biconformal current of the form
$\mathcal{J}=\frac{3}{2}J^{a}\mathbf{d}y_{a}-\frac{1}{2}y_{a}J_{\quad,b}%
^{a}\mathbf{d}x^{b}$, the biconformal field equation for $\mathcal{F}$ is
satisfied if and only if the original Minkowski space Yang-Mills equation is
satisfied. By construction, the Minkowski space Yang-Mills solution guarantees
that the biconformal equations are satisfied. Conversely, suppose we can write
the biconformal current as%
\begin{equation}
\mathcal{J}=\frac{3}{2}J^{a}\mathbf{d}y_{a}-\frac{1}{2}y_{a}J_{\quad,b}%
^{a}\mathbf{d}x^{b}%
\end{equation}
Then we need only the ansatz $\mathcal{A}=\mathbf{A}+\frac{1}{2}F^{ab}%
y_{a}\mathbf{d}y_{b}$ to find that the field equation takes the form
\begin{equation}
-\frac{1}{2}y_{a}F_{\quad,cb}^{ab}\mathbf{d}x^{c}+\frac{3}{2}F_{\quad,a}%
^{ab}\mathbf{d}y_{b}=\frac{3}{2}J^{a}\mathbf{d}y_{a}-\frac{1}{2}y_{a}%
J_{\quad,b}^{a}\mathbf{d}x^{b}%
\end{equation}
Equating the coefficients of $\mathbf{d}y_{b}$ gives%
\begin{equation}
F_{\quad,a}^{ab}=J^{b}%
\end{equation}
with the remaining coefficients of $\mathbf{d}x^{c}$ matching identically.
This completes the proof.

\section{Non-Abelian Yang-Mills in biconformal space}

In this section we generalize the results of Section 5, embedding a general
$4$-dimensional Yang-Mills theory into $8$-dimensional biconformal space. For
non-Abelian Yang-Mills theories, the calculations are similar to those given
in the previous section. We will state the result here and provide details in
the Appendix.

Consider the extension of a generic SUSY Yang-Mills theory to biconformal
space. We consider only the bosonic sector for simplicity. Beginning with the
$4$-dim equations%
\begin{align*}
\mathbf{A}  &  =\mathbf{A}^{B}G_{B}=A_{a}^{B}\left(  x\right)  \mathbf{d}%
x^{a}\\
\mathbf{F}  &  =\mathbf{F}^{B}G_{B}\\
&  =\mathbf{dA}+\mathbf{A}\wedge\mathbf{A}\\
&  =\mathbf{dA}^{B}G_{B}+\frac{1}{2}\mathbf{A}^{B}\wedge\mathbf{A}^{C}\left[
G_{B},G_{C}\right] \\
\mathbf{DF}  &  \equiv\mathbf{dF-F}\wedge\mathbf{A+A}\wedge\mathbf{F}=0
\end{align*}
where the covariant exterior derivative of a vector $\mathbf{v}^{A}$ in the
Lie algebra is given by%
\[
\mathbf{Dv}^{A}=\mathbf{dv}^{A}+c_{BC}^{\quad\quad A}\mathbf{A}^{B}%
\wedge\mathbf{v}^{C}%
\]
We extend by defining,%
\begin{align}
\mathcal{A}  &  =\mathbf{A}+\frac{1}{2}F^{Aab}y_{a}G_{A}\mathbf{d}y_{b}\\
\mathcal{F}  &  =\mathbf{d}\mathcal{A}+\mathcal{A}\wedge\mathcal{A}%
\end{align}
or, more explicitly,%
\begin{align}
\mathcal{A}^{B}  &  =\mathbf{A}^{B}+\frac{1}{2}F^{Bab}y_{a}\mathbf{d}%
y_{b}\label{Potential}\\
\mathcal{F}^{A}  &  =\mathbf{F}^{A}+\frac{1}{2}F^{Aab}\mathbf{d}%
y_{a}\mathbf{d}y_{b}-\frac{1}{2}y_{a}F_{\qquad,c}^{Aab}\mathbf{d}%
y_{b}\mathbf{d}x^{c}+\mathbf{A}^{B}\wedge\frac{1}{2}y_{c}F^{Ccd}%
\mathbf{d}y_{d}c_{BC}^{\quad A}\nonumber\\
&  +\frac{1}{8}c_{BC}^{\quad A}y_{a}F^{Bab}y_{c}F^{Ccd}\mathbf{d}y_{b}%
\wedge\mathbf{d}y_{d} \label{Field strength}%
\end{align}

If in $8$-dimensional biconformal space we use these fields to form the pure
Yang-Mills action with source,
\begin{equation}
S_{BC}=\int K_{AB}\left(  \mathcal{F}^{A\ast}\mathcal{F}^{B}+\mathcal{A}%
^{A\ast}\mathcal{J}^{B}\right)
\end{equation}
then the $8$-dimensional field equations take the form,%
\begin{equation}
^{\ast}\mathbf{D}^{\ast}\mathcal{F}^{A}=\mathcal{J}^{A}%
\end{equation}
where we define $\mathcal{J}^{A}$ to be the $8$-dim current,%
\begin{align}
\mathcal{J}^{A}  &  =\frac{1}{2}\left(  \left(  J^{Aa}{}_{;n}+\frac{5}%
{4}c_{GF}{}^{A}F^{Gap}F_{pn}^{F}\right)  y_{a}\mathbf{d}x^{n}\right.
\nonumber\\
&  +\left.  \left(  \frac{3}{2}J^{An}+\frac{1}{8}c_{BC}^{\quad A}y_{a}%
y_{c}\left(  -J^{Ba}F^{Ccn}-5F^{Ban}{}_{;d}F^{Ccd}\right)  \right)
\mathbf{d}y_{n}\right)  \label{biconformal current}%
\end{align}
By inspection, we see that if we set $y=0$, we return to the original $4$-dim
field equations. Furthermore, direct calculation shows that $\mathcal{J}^{A}$
is conserved if and only if the $4$-dim current, $J^{An}$, is conserved. That
is,
\begin{equation}
^{\ast}\mathbf{D}^{\ast\ast}\mathbf{D}^{\ast}\mathcal{F}^{A}=\frac{5}{4}%
J^{An}{}_{;n}%
\end{equation}
As a result, the $8$-dim field equations are satisfied if and only if the
$4$-dim equations hold. It is of some interest to note that the biconformal
current eq.(\ref{biconformal current}) depends on both the field and current
of the $4$-dim theory (see Appendix). This opens up the possibility that
vacuum solutions to the biconformal field theory could correspond to
Yang-Mills theories with sources -- that is, we have a geometric description
of certain matter fields.

This construction provides a new arena for investigating the behavior of such
standard $4$-dimensional theories as $SU(N)$ Yang-Mills or $N=4$ SUSY
Yang-Mills in 4-dimensions. As essentially the only known finite quantum field
theory, the study of $N=4$ SUSY Yang-Mills might provide insight into the
quantum behavior of gauge theories in biconformal space. Additionally,
although the $4$-dimensional, flat space, SUSY Yang-Mills theory is embedded
into another Ricci flat geometry in 8-dimensions, the geometry of biconformal
space is non-trivial. For example, the symplectic and phase space structures
of biconformal space \cite{QM} could possibly yield interesting new effects in
the extended Yang-Mills theory.

In summary of the previous two sections, we have shown that the co-tangent
bundle of Minkowski space extends to a biconformal bundle, and that any
Yang-Mills theory on $4$-dim Minkowski space correspondingly extends to an
equivalent field theory on a biconformal space. This result is important for
establishing that a Yang-Mills theory in biconformal space may be an
inherently $4$-dim theory. A similar result holds for biconformal gravity,
where it has been shown that the field equations for a curved biconformal
geometry reduce to general relativity on a $4$-dim submanifold \cite{JW}.
Understanding that $8$-dim biconformal spaces describe $4$-dim physics helps
clarify the naturalness of our final result below. We turn now to
$4$-dimensional Yang-Mills theories that lift to curved biconformal spaces.

\section{A special class of biconformal gravity solutions}

We now investigate how an unusual choice of gauge group in $4$-dimensions can
lead to a new class of biconformal Yang-Mills gravity solutions. Suppose we
begin with a flat space non-compact $SU(2,2|N)$ SUSY Yang-Mills theory on
$4$-dim Minkowski space$.$ As was demonstrated in the previous section, we can
write an equivalent field theory on biconformal space. However, the gauge
group of this Yang-Mills theory is identical to the gauge group of biconformal
supergravity \cite{AW}. This means that we can identify the Yang-Mills field
strengths with the curvatures of the biconformal Yang-Mills (super-) gravity theory.

The fact that makes this possible is that these field strengths, defined on an
$8$-dim manifold, satisfy the same conformal structure equations as the
$4$-dim theory. This means that we can identify the potentials $\mathcal{A}%
^{B}$ with the biconformal connection $\mathbf{\omega}^{A}$, and the
Yang-Mills curvatures $\mathcal{F}^{A}$ with the biconformal curvatures
$\mathbf{\Omega}^{A}$. Since we have chosen a Yang-Mills type action,
eq.(\ref{BYMG}), to describe the biconformal gravity theory, and used the
metric-independent volume form, the field equations are also identical.

The \textit{only} difference between the biconformally extended Yang-Mills
equations and the biconformal gravity equations is that (before imposing the
field equations) the biconformal curvatures are generic $2$-forms in $8$-dim,
while the extended Yang-Mills field strengths take the particular form above.
However, identifying the field strengths and curvatures is acceptable as an
ansatz for a biconformal solution. Allowing for differences of homothetic
gauge, $g,$ conformal gauge, $h,$ and coordinates, this ansatz takes the form:%
\begin{equation}
g\Omega^{A}g^{-1}\left(  u\left(  x,y\right)  ,v\left(  x,y\right)  \right)
=h\mathcal{F}^{A}h^{-1}\left(  x,y\right)  \label{Correspondence}%
\end{equation}
Since the extended Yang-Mills equations are solved if the original $4$-dim
Yang-Mills theory is satisfied, we have written a class of solutions to the
biconformal field equations. It is eq.(\ref{Correspondence}) that provides the
equivalence between a $4$-dim gauge theory on flat space and an $8$-dim
biconformal gravity theory.

Further, the gravity solutions are $4$-dimensional, because the full
curvatures are determined by the connection on the $4$-dim submanifold found
by holding $y$ constant. That is, there are two ways to regard eq.(\ref{Field
strength}). First, regarded as a Yang-Mills field strength, $\left.
\mathcal{F}^{A}\right|  _{y=y_{0}}$ reduces to a Yang-Mills field on $4$-dim
Minkowski space. Second, regarded as a biconformal curvature, $\left.
\mathbf{\Omega}^{A}\right|  _{y=y_{0}}=\left.  \mathcal{F}^{A}\right|
_{y=y_{0}}$reduces to a gravity solution on a $4$-dim submanifold of a
biconformal space. This is a correspondence between a theory containing
gravity and a gauge theory without it. It is worth stressing again here that
this relationship between flat and curved geometries is only possible because
of the symplectic structure of biconformal spaces. It is that structure that
allows us to write the action without introducing a metric, so that the
resulting field equations are identical to those of the flat space,
non-compact, SUSY Yang-Mills theory.

\section{Conclusions}

We have accomplished our goal of constructing a Yang-Mills theory of gravity
by using the biconformal gauging of the conformal group. In addition we have
provided a systematic extension of flat space Yang-Mills theories to
biconformal space. The resulting geometries are an arena for both biconformal
Yang-Mills gravity theory and extended Yang-Mills theories and possesses a
number of new and interesting properties:

\begin{enumerate}
\item Biconformal Yang-Mills gravity theories describe $4$-dim,
scale-invariant general relativity in the case of slowly changing fields.

\item The non-trivial systematic biconformal extension of any $4$-dim
Yang-Mills theory to a Yang-Mills theory on a $8$-dim biconformal space
provides a new arena for analyzing the properties and behavior of flat space
Yang-Mills theories.

\item By applying the biconformal extension above to a 4-dimensional pure
Yang-Mills theory with conformal symmetry, we establish a 1-1, onto mapping
between a set of gravitational gauge theories and $4$-dim, flat space gauge theories.

\item Since the biconformal Yang-Mills current, $\mathcal{J}^{A}$ depends on
both the $4$-dimensional fields and currents, some vacuum configurations of
the 8-dim field strength give rise to a non-trivial source term in the $4$-dim
Yang-Mills theory. This provides a geometric origin for some matter fields.
\end{enumerate}

The extension in (2.) above is obtained by extending the theory from a $4$-dim
(flat) manifold to its $8$-dim co-tangent bundle ($T_{\ast}M$), then requiring
a conformal connection, $\mathbf{\omega}^{\Sigma}$, on $T_{\ast}M$ to obtain a
Yang-Mills theory on a biconformal space. Although the extension is
non-trivial, it is readily invertible. After extension, the original theory
may be re-obtained by setting the $y$-coordinates to zero.

In the special class of Yang-Mills gravity solutions described in (3.), we
choose simple but non-compact $O\left(  4,2\right)  $ or $SU\left(
2,2\right)  $ Yang-Mills theories, and for their supersymmetric extension,
$SU(2,2|N)$ non-compact Yang-Mills theory. Then, biconformal Yang-Mills
gravity theory has a class of solutions for the curvatures, $\Omega^{A}$, that
satisfy $\Omega^{A}=\mathcal{F}^{A}$, where $\mathcal{F}^{A}$ is an extended
$SU(2,2|N)$ Yang-Mills field strength. These curvatures, $\Omega^{A},$ satisfy
the gravitational field equations if and only if the original Yang-Mills field
strengths, $F^{A}$, satisfy the super Yang-Mills equations in $4$-dim
Minkowski space. The fact that the biconformal space described in (3.) is a
curved geometry, while the equivalent Yang-Mills system has Minkowski base
manifold is made possible by the symplectic structure of biconformal spaces
and their resulting dimensionless volume forms.

The field equations for the SUSY Yang-Mills theory in $4$-dim are completely
determined by the potentials $A_{\mu}^{\Sigma}$ $(x)$ and as a result, contain
a total of $60$ degrees of freedom. When it is possible to impose both Dirac
and Coulomb type gauge conditions, there will remain $30$ degrees of freedom,
two for each potential. When it is possible to impose both Dirac and Coulomb
type gauge conditions, there will remain $30$ degrees of freedom, two for each
potential. However, the torsion-free biconformal field equations are
completely determined by the solder form, $e^{a}(x)$, which, subtracting $7$
for the gauge choice, is a total of only nine degrees of freedom. As a result,
there is vastly more freedom in the class of extended Yang-Mills solutions to
biconformal gravity than in the torsion-free solutions. Moreover, the
condition of vanishing torsion is not gauge invariant in the Yang-Mills theory
(though it is in biconformal space). Allowing for homothetic $\left(
g\right)  $ and coordinate freedom in the biconformal solutions, and the
conformal symmetry of the Yang-Mills solution $\left(  h\right)  $ in equating
the fields
\[
\left(  g\Omega^{A}g^{-1}\left(  u\left(  x,y\right)  ,v\left(  x,y\right)
\right)  =h\mathcal{F}^{A}h^{-1}\left(  x,y\right)  \right)
\]
we make the conjecture that all torsion-free solutions to the biconformal
field equations can be identified with extended Yang-Mills theories.

If this conjecture holds, then there is a 1-1 correspondence between solutions
for torsion-free biconformal supergravity with a class of $4$-dim non-compact
SUSY Yang-Mills theories. In any case, we have shown that the weak-field limit
of extended Yang-Mills solutions gives rise to a gravity theory on a $4$-dim
submanifold of biconformal space. Further study is required to know if this
set includes torsion free solutions \cite{Spencer}.

While the theories presented in this work have been purely classical, the
quantum properties of the biconformal gravity theory are a potentially rich
topic for further investigation. Since the field equations obtained from the
Yang-Mills gravity actions are linear in the curvatures, we do not expect
ghosts for the reasons that are common in standard conformal gravity theories.
However, because the theory is formulated in a $8$-dim symplectic space, it
not clear how to apply the standard techniques of ghost analysis. As a result,
the existence of ghosts in generic biconformal supergravity is a topic for
further study.

\section{Appendix: Explicit biconformal extension of non-Abelian Yang-Mills theory}

Here we consider the biconformal extension of a non-Abelian Yang-Mills theory.
Suppose we have a gauge theory with Lie gauge group $\mathcal{G}$, (group
elements $g,$ generators $G_{A},$ structure constants $c_{AB}{}^{C}$) in
$4$-dim Minkowski space, with potential, field strength and Bianchi identity,
\begin{align}
\mathbf{A}  &  =\mathbf{A}^{B}G_{B}=A_{a}^{B}\left(  x\right)  \mathbf{d}%
x^{a}\nonumber\\
\mathbf{F}  &  =\mathbf{F}^{B}G_{B}\nonumber\\
&  =\mathbf{dA}+\mathbf{A}\wedge\mathbf{A}\nonumber\\
&  =\mathbf{dA}^{B}G_{B}+\frac{1}{2}\mathbf{A}^{B}\wedge\mathbf{A}^{C}\left[
G_{B},G_{C}\right] \nonumber\\
\mathbf{DF}  &  \equiv\mathbf{dF-F}\wedge\mathbf{A+A}\wedge\mathbf{F}=0
\end{align}
In components, for any element of the Lie algebra, we have
\begin{equation}
\mathbf{Dv}^{A}=\mathbf{dv}^{A}+c_{BC}^{\quad\quad A}\mathbf{A}^{B}%
\wedge\mathbf{v}^{C}%
\end{equation}
With these fields, the standard Yang-Mills action with source%
\begin{equation}
S_{M}=\int K_{AB}\left(  \mathbf{F}^{A\ast}\mathbf{F}^{B}+\mathbf{A}^{A\ast
}\mathbf{J}^{B}\right)
\end{equation}
leads to the field equation the $4$-dim equation,
\begin{align}
^{\ast}\mathbf{D}^{\ast}\mathbf{F}  &  =-\mathbf{J}\\
\mathcal{D}_{c}F_{\quad e}^{Ac}  &  =-J_{\quad e}^{A}\nonumber
\end{align}
Where
\begin{equation}
\mathcal{D}_{a}F_{\quad e}^{Ac}=F_{\quad e,a}^{Ac}+c_{BC}^{\quad\quad
A}F_{\quad e}^{Cc}A_{a}^{B}%
\end{equation}

We wish to extend this to biconformal space. As before, we construct the
biconformal space from the co-tangent bundle with coordinates $\left(
x^{a},y_{b}\right)  $ extended by $7$-dim homothetic fibers. We may then
define a biconformal connection on the full $15$-dim bundle,
\begin{align}
\mathbf{\omega}  &  =-y_{a}\mathbf{d}x^{a}\nonumber\\
\mathbf{\omega}^{a}  &  =\mathbf{d}x^{a}\nonumber\\
\mathbf{\omega}_{a}  &  =\mathbf{d}y_{a}-\alpha\left(  y_{a}y_{b}-\frac{1}%
{2}y^{2}\eta_{ab}\right)  \mathbf{d}x^{b}\nonumber\\
\mathbf{\omega}_{b}^{a}  &  =\left(  y_{b}\mathbf{d}x^{a}-y^{a}\mathbf{d}%
x_{b}\right)
\end{align}
satisfying
\begin{align}
\mathbf{d\omega}  &  =\mathbf{\omega}^{a}\mathbf{\omega}_{a}\nonumber\\
\mathbf{d\omega}^{a}  &  =\mathbf{\omega}^{b}\mathbf{\omega}_{b}%
^{a}+\mathbf{\omega\omega}^{a}\nonumber\\
\mathbf{d\omega}_{a}  &  =\mathbf{\omega}_{a}^{b}\mathbf{\omega}%
_{b}+\mathbf{\omega}_{a}\mathbf{\omega}\nonumber\\
\mathbf{d\omega}_{b}^{a}  &  =\mathbf{\omega}_{b}^{c}\mathbf{\omega}_{c}%
^{a}+2\Delta_{db}^{ac}\mathbf{\omega}_{c}\mathbf{\omega}^{d}%
\end{align}
If the Yang-Mills group is the superconformal group, $SU\left(  2,2|N\right)
$, we identify the extended Yang-Mills potentials with the biconformal connection.

In either case, we define the extended Yang-Mills potential,
\begin{align}
\mathcal{A}  &  =\mathbf{A}+\frac{1}{2}F^{Aab}y_{a}G_{A}\mathbf{d}y_{b}\\
\mathcal{A}^{B}  &  =\mathbf{A}^{B}+\frac{1}{2}F^{Bab}y_{a}\mathbf{d}y_{b}%
\end{align}
The potential remains a Lie algebra valued $1$-form, and therefore satisfies
the usual Maurer-Cartan structure equations. It follows that the field is
given by
\begin{align}
\mathcal{F}  &  =\mathbf{d}\mathcal{A}+\mathcal{A}\wedge\mathcal{A}\nonumber\\
\mathcal{F}^{A}  &  =\mathbf{F}^{A}+\frac{1}{2}F^{Aab}\mathbf{d}%
y_{a}\mathbf{d}y_{b}-\frac{1}{2}y_{a}F_{\qquad,c}^{Aab}\mathbf{d}%
y_{b}\mathbf{d}x^{c}\nonumber\\
&  +\mathbf{A}^{B}\wedge\frac{1}{2}y_{c}F^{Ccd}\mathbf{d}y_{d}c_{BC}^{\quad
A}+\frac{1}{8}c_{BC}^{\quad A}y_{a}F^{Bab}y_{c}F^{Ccd}\mathbf{d}y_{b}%
\wedge\mathbf{d}y_{d}%
\end{align}
As in the $U(1)$ case, the biconformal potential and field reduce to the 4-dim
potential and field when $y=0$.

In general, we may think of the extension from $\mathbf{A}$ to $\mathcal{A}$
as a change of connection, so that the difference is a tensor. The covariant
derivative then will change by the addition of a tensorial term, and the
curvature by the addition of the covariant derivative of the new tensor, plus
the square of the new tensor. The calculation above confirms this.

The Bianchi identity also generalizes to
\[
\mathbf{D}\mathcal{F}=\left(  \mathbf{d}\mathcal{F}^{A}+c_{BC}^{\qquad
A}\mathcal{A}^{B}\wedge\mathcal{F}^{C}\right)  G_{A}=0
\]
where the covariant derivative is given by%
\begin{equation}
\mathbf{D}=\mathcal{D}_{p}\mathbf{d}x^{p}+\mathcal{D}^{p}\mathbf{d}y_{p}%
\end{equation}
with%
\begin{align}
\mathcal{D}_{p}X^{A}  &  =\partial_{p}X^{A}+c_{EF}{}^{A}A^{E}X^{F}\nonumber\\
\mathcal{D}^{p}X^{A}  &  =\partial^{p}X^{A}+\frac{1}{2}c_{EF}{}^{A}%
F^{Elp}X^{F}y_{l}%
\end{align}

The field equations from the action written with the $8$-dim fields
\begin{equation}
S_{BC}=\int K_{AB}\left(  \mathcal{F}^{A\ast}\mathcal{F}^{B}+\mathcal{A}%
^{A\ast}\mathcal{J}^{B}\right)
\end{equation}
then take the form,%
\begin{equation}
^{\ast}\mathbf{D}^{\ast}\mathcal{F}^{A}=\mathcal{J}^{A}%
\end{equation}
where direct computation yields%
\begin{align}
^{\ast}\mathbf{D}^{\ast}\mathcal{F}^{A}  &  =\frac{1}{2}\left(  \left(
F^{Apa}{}_{;pn}+\frac{5}{4}c_{GF}{}^{A}F^{Gap}F_{pn}^{F}\right)
y_{a}\mathbf{d}x^{n}\right. \\
&  \left.  +\left(  \frac{3}{2}F^{Apn}{}_{;p}+\frac{1}{8}c_{BC}^{\quad A}%
y_{a}y_{c}\left(  -F^{Bpa}{}_{;p}F^{Ccn}-5F^{Ban}{}_{;d}F^{Ccd}\right)
\right)  \mathbf{d}y_{n}\right) \nonumber
\end{align}
and $\mathcal{J}^{A}$ is the $8$-dim current,%
\begin{align}
\mathcal{J}^{A}  &  =\frac{1}{2}\left(  \left(  J^{Aa}{}_{;n}+\frac{5}%
{4}c_{GF}{}^{A}F^{Gap}F_{pn}^{F}\right)  y_{a}\mathbf{d}x^{n}\right.
\nonumber\\
&  +\left.  \left(  \frac{3}{2}J^{An}+\frac{1}{8}c_{BC}^{\quad A}y_{a}%
y_{c}\left(  -J^{Ba}F^{Ccn}-5F^{Ban}{}_{;d}F^{Ccd}\right)  \right)
\mathbf{d}y_{n}\right)
\end{align}
It is evident that if we set $y=0$, we return to the original $4$-dim field equations.

\bigskip Furthermore, the $8$-dim current, $\mathcal{J}^{A}$ is conserved if
and only if the $4$-dim current, $J^{An}$, is conserved. That is,
\begin{equation}
^{\ast}\mathbf{D}^{\ast\ast}\mathbf{D}^{\ast}\mathcal{F}^{A}=\frac{5}{4}%
J^{An}{}_{;n}%
\end{equation}
As a result, the $8$-dim field equations are satisfied if and only if the
$4$-dim equations hold.

\end{document}